\documentstyle [eqsecnum,preprint,aps,epsf]{revtex}

\makeatletter
\def\footnotesize{\@setsize\footnotesize{10.0pt}\xpt\@xpt
\abovedisplayskip 10\p@ plus2\p@ minus5\p@
\belowdisplayskip \abovedisplayskip
\abovedisplayshortskip  \z@ plus3\p@
\belowdisplayshortskip  6\p@ plus3\p@ minus3\p@
\def\@listi{\leftmargin\leftmargini
\topsep 6\p@ plus2\p@ minus2\p@\parsep 3\p@ plus2\p@ minus\p@
\itemsep \parsep}}
\long\def\@makefntext#1{\parindent 5pt\hsize\columnwidth\parskip0pt\relax
\def\strut{\vrule width0pt height0pt depth1.75pt\relax}%
$\m@th^{\@thefnmark}$#1}
\long\def\@makecaption#1#2{%
\setbox\@testboxa\hbox{\outertabfalse %
\reset@font\footnotesize\rm#1\penalty10000\hskip.5em plus.2em\ignorespaces#2}%
\setbox\@testboxb\vbox{\hsize\@capwidth
\ifdim\wd\@testboxa<\hsize %
\hbox to\hsize{\hfil\box\@testboxa\hfil}%
\else %
\footnotesize
\parindent \ifpreprintsty 1.5em \else 1em \fi
\unhbox\@testboxa\par
\fi
}%
\box\@testboxb
} %
\global\firstfigfalse
\global\firsttabfalse
\def\tabular{\let\@halignto\@empty\@tabular}
\def\endtabular{\crcr\egroup\egroup $\egroup}
\expandafter \def\csname tabular*\endcsname #1{\def\@halignto{to#1}\@tabular}
\expandafter \let \csname endtabular*\endcsname = \endtabular
\def\@tabular{\leavevmode \hbox \bgroup $\let\@acol\@tabacol
   \let\@classz\@tabclassz
   \let\@classiv\@tabclassiv \let\\\@tabularcr\@tabarray}
\def\endtable{%
\global\tableonfalse\global\outertabfalse
{\let\protect\relax\small\vskip2pt\@tablenotes\par}\xdef\@tablenotes{}%
\egroup
}%
\global\firstfigfalse
\global\firsttabfalse

\def\undertilde{\mathpalette\utild@}%
\newbox\utildbox
\def\utild@#1#2%
    {%
    \setbox0=\hbox {$\m@th #1 {#2}$}%
    \dimen@=\dp0\advance\dimen@ by 0.2 ex\dp0=\dimen@%
    \setbox\utildbox=\hbox {$\tilde {\hphantom {\copy0}}$}%
    \setbox0=\vtop {\offinterlineskip\box0\box\utildbox}%
    \dp0=0.4ex\box0\relax%
    }%

\makeatother

\def\eff{_{\rm eff}}
\def\bare{_{\rm bare}}
\def\dlam{\delta\lambda\eff}
\def\F{{\cal F}}
\def\intFF{\int_{pq} \F_{pq}}
\def\v#1{{\bf #1}}
\def\countera{q}
\def\counterb{r}

\begin {document}


\preprint {UW/PT-96-27}

\title  {The tricritical point of finite-temperature phase transitions
  in large $N$(Higgs) gauge theories.}

\author {Peter Arnold and David Wright}

\address
    {%
    Department of Physics,
    University of Washington,
    Seattle, Washington 98195
    }%
\date {October 1996}

\maketitle

\begin {abstract}%
{%
{%
Gauge theories broken by a single Higgs
field are known to have first-order phase transitions in temperature
if $\lambda/g^2 \ll 1$, where $g$ is the gauge coupling
and $\lambda$ the Higgs self-coupling.  If the theory is extended from one
to $N$ Higgs doublets,
with U($N$) flavor symmetry, the transition is known to be
second order for $\lambda/g^2 \gtrsim 1$ in the $N\to\infty$ limit.  We show
that one can in principal compute the tricritical value of $\lambda/g^2$,
separating first from second-order transitions, to any order in $1/N$.  In
particular, scalar fluctuations at the transition damp away the usual problems
with the infrared behavior of high-temperature non-Abelian gauge theories.  We
explicitly compute the tricritical value of $\lambda/g^2$ for U(1)
and SU(2) gauge theory to next-to-leading order in $1/N$.
}%
\ifpreprintsty
\thispagestyle {empty}
\newpage
\thispagestyle {empty}
\vbox to \vsize
    {%
    \vfill \baselineskip .28cm \par \font\tinyrm=cmr7 \tinyrm \noindent
    \narrower
    This report was prepared as an account of work sponsored by the
    United States Government.
    Neither the United States nor the United States Department of Energy,
    nor any of their employees, nor any of their contractors,
    subcontractors, or their employees, makes any warranty,
    express or implied, or assumes any legal liability or
    responsibility for the product or process disclosed,
    or represents that its use would not infringe privately-owned rights.
    By acceptance of this article, the publisher and/or recipient
    acknowledges the U.S.~Government's right to retain a non-exclusive,
    royalty-free license in and to any copyright covering this paper.%
    }%
\fi
}%
\end {abstract}
\newpage


\section {Introduction}

   The possibility of explaining the observed baryon number of the universe by
physics occurring at the electroweak phase transition has, in recent years,
renewed interest in understanding how to assess the existence, order, and
strength of phase transitions in gauge theories, such as electroweak theory,
where gauge bosons get mass by the Higgs mechanism.
The phase structure of
the electroweak sector of the minimal standard model, with a single Higgs
doublet, makes a good starting point for exploring such transitions.  It has
long been appreciated
\cite{Kirzhnits&Linde}
that the phase transition is first-order in the
limit that the zero-temperature
Higgs boson mass is small compared to the W boson mass,
{\it i.e.} when
$\lambda \ll g^2$, where $g$ is the electroweak gauge
coupling and $\lambda$ is the Higgs self-coupling.  In this limit, a
perturbative analysis of the phase transition is adequate to establish its
order and compute its physical properties.  (Here and throughout, we assume
$\lambda$ and $g^2$ are both small.)  What has been more difficult is to study
the transition when $\lambda \gtrsim g^2$.  In this limit, perturbation theory
breaks down due to large infrared fluctuations characteristic of critical or
near-critical behavior.  One response is to turn to numerical simulations of
the transition.  It is interesting, however, to see what can be said about the
transition analytically if one modifies the theory to make it more tractable.
For example, if the three spatial dimensions are replaced by $4{-}\epsilon$
dimensions, where $\epsilon\ll 1$, then it is known that the transition
remains first-order for any finite $\lambda/g^2$
\cite{Chen&Lubensky&Nelson,Arnold&Yaffe}.
If the Higgs sector is
generalized to contain $N$ Higgs doublets with U($N$) symmetry, then in the
$N\to\infty$ limit the transition is first-order for $\lambda/g^2 \ll 1/N$
and second-order for $\lambda/g^2 \gg 1/N$
\cite{Arnold&Yaffe}.
Recent numerical simulations
for $N{=}1$, in contrast, suggest that the first-order transitions end at a
critical value of $\lambda/g^2$ above which there is no phase transition
whatsoever \cite{Kajantie}.%
\footnote{
   We should emphasize that one of the pieces of evidence presented in
   ref.~\cite{Kajantie}---the large volume dependence of the $\Phi^\dagger\Phi$
   susceptibility---has a loophole which is realized in a simple and
   relevant example.
   Ref.~\cite{Kajantie} shows evidence that this
   susceptibility approaches a constant
   in the large volume limit, for Higgs masses above roughly 80 GeV.
   In a second-order transition, the large volume behavior of this
   susceptibility should be an analytic function of $V^{-1}$ plus
   a non-analytic scaling piece whose leading term is $V^{\alpha/3\nu}$,
   where $\alpha$ and $\nu$ are respectively the specific heat and
   correlation length exponents.  However, there are systems with
   second-order transitions where $\alpha/\nu$ is {\it negative}, and
   so the measured susceptibility would indeed approach a constant
   rather than diverge.  A relevant example is the pure scalar sector
   of electroweak theory itself, the O(4) model, where
   $\alpha/3\nu = -0.33(4)$ (see table~\ref{tab:o4}).  A more convincing
   demonstration of the absence of a transition is the plot in
   ref.~\cite{Kajantie} of the inverse correlation length
   vs.\ temperature, which
   shows no suggestion of a divergence in the correlation length.
}

   The goal of the present work is to extend understanding of the large
$N$ limit beyond leading order in $1/N$, studying in particular the
critical value of $\lambda/g^2$ demarking the end of first-order
transitions.  We emphasize that large $N$ here refers to the number of
scalar fields and not to the replacement of the gauge group by SU($N$).

   Usually, studying critical behavior of weakly coupled field theories
is {\it more} difficult than studying those theories far from the transition,
because long-distance fluctuations appear at the transition whose
physics is non-perturbative.
Small $N$ pure scalar theories, for example, are easy to study far from
the transition but difficult near the transition for this reason.
Amusingly, large $N$(scalar) non-Abelian gauge theories are exactly
the opposite:
At temperatures far above the transition,
electric forces are Debye screened in the hot plasma but magnetic forces
are not, and the non-Abelian nature of the forces gives rise to magnetic
confinement at large distances, which cannot be treated perturbatively.
But, as we shall discuss, {\it at} the phase transition
long-distance scalar fluctuations screen the magnetic forces
sufficiently to prevent magnetic confinement.
The long-distance scalar fluctuations themselves are treatable in
a $1/N$ expansion just as in large $N$ pure scalar theories.

In the remainder of this introduction, we briefly discuss how large one
might suspect $N$ has to be for the large $N$ expansion to be useful.
Then we discuss whether a moderately large $N$ Higgs sector is
phenomenologically viable.
In section 2, we will fix notation and briefly review that the
problem of finite temperature phase transitions in 3+1 dimensions
is equivalent to the study of field theories in 3 Euclidean dimensions.
Then we discuss the power counting of the loop expansion and why
magnetic confinement is not a problem at the transition.  We will also
see that the calculation of the tricritical value of $\lambda/g^2$
order by order in $1/N$ is conceptually more straightforward than
the calculation of many other
quantities.  In section 3, we carry out this computation to
next-to-leading order for the U(1) gauge theory.  Section 4 is devoted
to clearing up some minor subtleties of regularization of diagrams.
Finally, we carry out the next-to-leading order computation for SU(2) theory
in section 5.
same for SU(2) theory in section 5.

\subsection{How large in large $N$?}

If gauge interactions are ignored,
the scalar sector of the minimal standard model with a {\it single}
Higgs doublet is equivalent to an O(4) theory of four real scalar
fields and has a second-order transition.
Table~\ref{tab:o4} demonstrates
large $N$ results applied to this case ($g{=}0$).
Amusingly, next-to-leading order in $1/N$
actually gets in the right ballpark for various critical exponents.
Sadly,
this happy circumstance that 1 is just on the verge of being a large number of
doublets will not survive the inclusion of gauge interactions.

\begin {figure}
\begin {table}
\begin {center}
\tabcolsep 10pt

\def\p{\phantom{-}}
\begin {tabular}{|rlll|ll|}
\hline
            & \multicolumn{3}{c|}{Large $N$}    & \multicolumn{2}{c|}{actual}
\\
            & LO       & NLO      & NNLO        & \multicolumn{1}{|c}{series}
                                                             & monte carlo
\\
\hline
  $\gamma=$ & \p2      & \p1.392  &  1.188      & \p1.44(4)  & \p1.477(18)
\\
  $\nu=$    & \p1      & \p0.730  &  0.612      & \p0.73(2)  & \p0.7479(90)
\\
  $\beta=$  & \p0.5    & \p0.399  &  0.325      & \p0.38(1)  & \p0.3836(46)
\\
  $\delta=$ & \p5      & \p4.594  &  4.695      & \p4.82(5)  & \p4.851(22)
\\
  $\eta=$   & \p0.0675 & \p0.0554 &  0.0260     & \p0.03(1)  & \p0.0254(38)
\\
  $\alpha=$ & -1       & -0.189   &  0.163      & -0.19(6)   & -0.244(27)
\\
\hline
\end {tabular}
\end {center}
\caption
    {%
    Large $N$ expansion results \protect\cite{o4 large n}
    for critical exponents in the pure scalar
    case ($g$=0) for one Higgs doublet: the O(4) model.  Results are given
    for leading (LO), next to leading (NLO), and next to next to
    leading (NNLO) order in $1/N$.  The actual values, as predicted from
    series analysis
    ({\protect\cite{Nickel}} as cited in \protect\cite{Wilczek})
    or measured by monte carlo \protect\cite{Kanaya&Kaya}, are also shown.
    There are various scaling relationships between these quantities,
    so that only two of the above exponents are independent.
    }
\label {tab:o4}
\end {table}
\end {figure}

   At another extreme, one can analyze the phase structure for arbitrary
$N$ in $4-\epsilon$ spatial dimensions
\cite{Halperin&Lubensky&Ma,Arnold&Yaffe}.
One finds that the qualitative
picture given by the large $N$ limit---that there is a tricritical value
of $\lambda/g^2$ above which the transition is second order---is correct
when
\begin {eqnarray}
   N &>& 182.95 - 320.50 \epsilon + O(\epsilon^2) \,,
      \qquad\hbox{for U(1) with $N$ charged scalars} \,,
\\
   N &>& 359 - 495.4\epsilon + O(\epsilon^2) \,,
      \phantom{182.}\qquad\hbox{for SU(2) with $N$ scalar doublets} \,.
\end {eqnarray}
So, near four spatial dimensions, large $N$ means $N \gg 183$ or
$N \gg 359$, respectively!

A result of our calculation of the
tricritical point will be that
quantitative success of large $N$ in three dimensions appears to require
$N \gg 4$ charged scalars for U(1) theory and $N \gg 20$ doublets
for SU(2) theory.


\subsection{Is moderately large $N$ phenomenologically viable?}

Generalizing the one Higgs model to a $U(N)$ Higgs model is motivated
solely by the desire to find a theory whose phase transition is
analytically tractable.  Nonetheless, it's interesting to briefly
consider (just for fun!) whether such a model might be consistent with
real world phenomenology.

A $U(N)$ flavor-symmetric Higgs sector is a phenomenological disaster
because electroweak symmetry breaking will break the global $U(N)$
and produce massless Goldstone bosons.  The essential difference
between Higgs models susceptible to large $N$(scalar) analysis and generic
multiple Higgs models is the necessity of a large $N$ global symmetry.
There is no reason, however, that this symmetry need be continuous.
Though we do not study it in this paper, one could instead have a
Higgs sector with a {\it discrete} $N$-flavor permutation symmetry:%
\footnote{
  The purely scalar sector is a simple generalization of the cubic
  anisotropy model.  See, for example, ref.~\cite{cubic anisotropy}.
}
\begin {eqnarray}
   {\cal L} &\sim&
     |D\vec\Phi|^2
     + m^2 |\vec\Phi|^2
     + \lambda_1 |\vec\Phi|^4
     + \lambda_2 \sum_i (\Phi_i^* \Phi_i)^2
\nonumber\\ && \qquad
     + g_u \bar Q_{\rm R} (\Phi_1 + \cdots + \Phi_N) u_{\rm L} 
     + g_d \bar Q_{\rm R} \tau_2 (\Phi_1 + \cdots + \Phi_N)^* d_{\rm L}
   \,.
\end {eqnarray}
As an added bonus, the permutation symmetry prevents the tree-level
flavor-changing neutral currents that plague generic multiple scalar models.

Because the discrete flavor symmetry is spontaneously broken, the model
suggested above will produce cosmological domain walls which overclose
the universe.  This problem could by solved by the introduction of a
very small symmetry breaking term (which is natural in the sense of
't Hooft \cite{tHooft})
that would cause the domain walls to coalesce after they
were formed.

Some sort of $N>1$ models therefore seem acceptable phenomenologically.
It is worth noting that there is an important qualitative difference
between the $N=1$ and $N>1$ cases.  For $N>1$ there is, by construction,
a {\it global} flavor symmetry that is spontaneously broken at the electroweak
scale.  Such models therefore always have some sort of phase transition
(ignoring the tiny symmetry breaking term discussed above).
The $N=1$ model, in contrast, need not have a transition because,
technically, local symmetries are never spontaneously broken due to
Elitzur's theorem \cite{elitzur,Kajantie}.

We shall not analyze in any detail just how large $N$ could be in a
realistic theory except to make one, trivial observation: there is
a simple constraint from triviality.
Non-perturbative continuum scalar theories are not well defined,
and the effective strength of scalar
interactions is $O(N\lambda)$ instead of $O(\lambda)$.
The largest Higgs mass for which the theory can be sensible as an
effective theory therefore decreases roughly as $1/\sqrt N$ from the
$N{=}1$ limit of $O$(1~TeV).  $N \gg 100$ is clearly out of the picture.
On a related note, $N > 20$ would destroy the asymptotic
freedom of the SU(2) electroweak interactions in the standard model.

One can imagine that $N$ might be big enough for the large $N$ approximation
to be reasonable, but not too big to run into phenomenological problems.
There's no good reason, of course, why nature would
choose to be so peculiar.


\section {Large $N$ counting}

   The problem of studying a second-order (or very weakly first-order)
phase transition in weakly coupled quantum field theory, as one varies the
temperature, can be reduced to the problem of studying the phase transition in
three-dimensional Euclidean field theory, as one varies a mass.  In our
case, the three-dimensional theory is of the form%
\begin {equation}
   S = \int d^3 x \left\{ |D\vec\Phi|^2
           + {\textstyle{1\over4}} F^2
           + m^2 |\vec\Phi|^2
           + {\textstyle{1\over6}} \lambda |\vec\Phi|^4
           + {\textstyle{1\over90}} \eta |\vec\Phi|^6 \right\}\,,
\label {eq:S3}
\end {equation}
and the transition occurs as $m^2$ is varied through zero.  Let's take
a moment to briefly review this correspondence.

   Critical or near-critical behavior of phase transitions is governed
by the physics of long distances, which in our case is classical.
It is classical because the Bose density per mode, $1/(e^{-\beta E} -1)$,
becomes large for small energies $E$.  So one may study the 
long-distance, equilibrium
properties of such transitions by studying the classical statistical
mechanics of field theory in three spatial dimensions.  More formally, one
may start with the Euclidean formulation of finite-temperature quantum
field theory and then integrate out the physics of the small, periodic
Euclidean time direction.
(See refs.~\cite{3D reviews} for a review.)
Taking the additional step of integrating out the Debye-screened $A_0$
field,
one obtains an effective three-dimensional
theory of the form (\ref{eq:S3}) plus irrelevant interactions.
The parameters of the effective
three-dimensional theory can be perturbatively related to those
of the original 3+1 dimensional theory if the couplings are small.
One finds, up to higher-order corrections, that
\begin {equation}
   g^2 = g_4^2(T) \, T \,, \qquad
   \lambda = \lambda_4(T) \, T \,, \qquad
   \eta = O(g_4^6, \lambda^3) \,, \qquad
\label {eq:matching}
\end {equation}
where $g_4(T)$ and $\lambda_4(T)$ are the dimensionless couplings of the
3+1 dimensional theory at a renormalization scale of order the
temperature $T$.
One also finds that increasing the temperature through the transition
corresponds to varying the scalar mass $m$ in the effective theory
from $m^2 < 0$ to $m^2 > 0$.  So the problem of understanding the
phase transition of the original theory in temperature is equivalent
to understanding the phase transition of a three-dimensional theory
in $m^2$.  In the limit that the original couplings
$g_4^2$ and $\lambda_4$ are considered arbitrarily small, the problem of
finding the tricritical value of $\lambda_4/g_4^2$ in the original
theory is the same as finding the tricritical value of $\lambda/g^2$
in the three-dimensional theory.

   So focus on the three-dimensional theory (\ref{eq:S3}), and note that
$\lambda$ and $g^2$ have dimensions of mass.  Now consider the naive
perturbative expansion for some physical observable associated with a small
momentum scale $p$ and suppose we are in the symmetric phase $m^2 > 0$
and that $p \ll m$.  Then the scalars will decouple, and we must focus
on the non-Abelian interactions of the magnetic gauge fields.
(As mentioned earlier, the electric ones are Debye screened.)
By dimensional analysis, the loop expansion parameter for gauge interactions
is then $g^2/p$, and perturbation theory will fail once we try to explore
momentum scales $p \lesssim g^2$.
This is the source of the infrared problem for non-Abelian gauge
theories at high temperature.

\begin {figure}
\vbox
   {%
   \begin {center}
      \leavevmode
      
      \epsfbox [150 350 500 480] {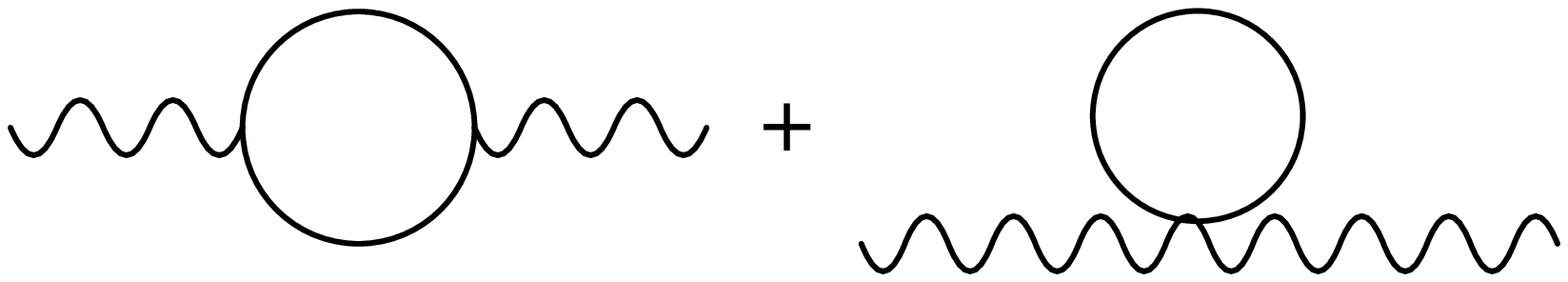}
   \end {center}
   \caption
       {%
       Leading contribution to the gauge boson self-energy.
       \label{fig:pi}
       }%
   }%
\end {figure}

\begin {figure}
\vbox
   {%
   \begin {center}
      \leavevmode
      
      \epsfbox [150 290 500 430] {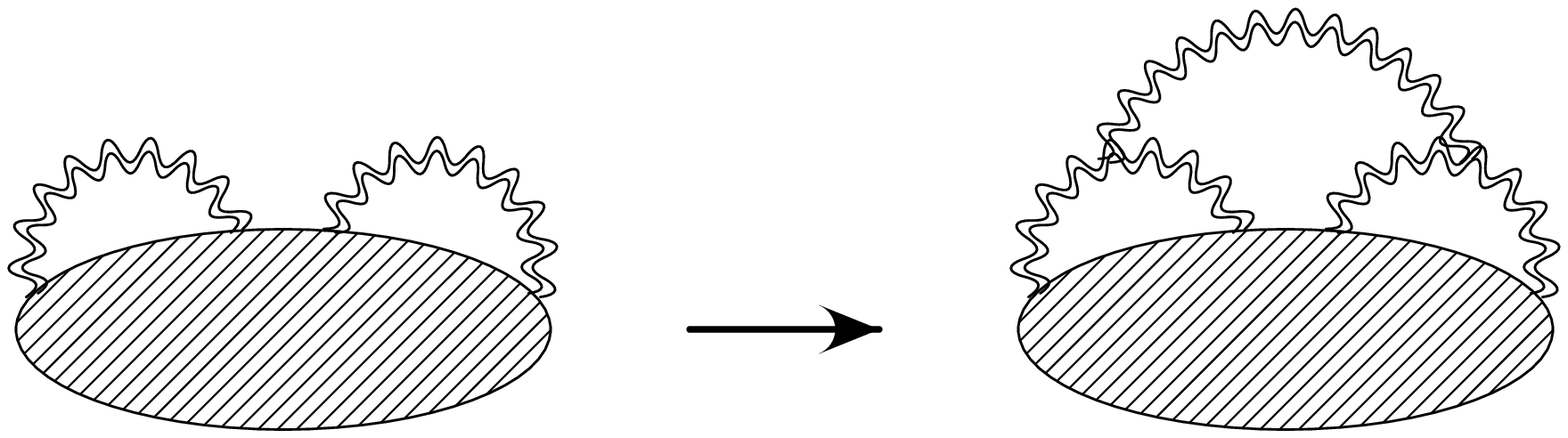}
   \end {center}
   \caption
       {%
       Adding a pair of non-Abelian interactions to a graph.
       \label{fig:add loop}
       }%
   }%
\end {figure}

   Now consider the case right at a second-order transition, where the scalar
mass is zero, and consider the effect of the self-energy diagrams of
fig.~\ref{fig:pi} on the gauge boson propagator.  By dimensional analysis, this
self-energy is
\begin {equation}
   \Pi(p) =  {a N g^2 p} \,,
\end {equation}
where $a$ is a numerical constant, and the gauge propagator becomes
\begin {equation}
   G(p) \sim {1 \over p^2 + a N g^2 p} \quad\to\quad {1 \over a N g^2 p}
   \quad \hbox{as $p\to0$} \,.
\label {eq:G}
\end {equation}
This is less divergent in the infrared than the perturbative propagator
$1/p^2$.  For distances $r \gg 1/Ng^2$, the scalar degrees of freedom have
screened the gauge propagator from $1/r$ behavior to $1/r^2$ behavior.  The
problematical interactions above the phase transition were non-Abelian gauge
interactions.  Now, with the propagator (\ref{eq:G}), such interactions are
under perturbative control for large $N$.  Loops of gauge bosons will in
general be infrared convergent, and the scale of the loop momenta will be $O(N
g^2)$ if the external momentum is small.  The cost of adding a new pair of
non-Abelian interactions to a graph, such as depicted in
fig.~\ref{fig:add loop}, is then, by dimensional analysis, generically
\begin {equation}
   g^2 \times {1\over N g^2} \sim {1\over N} \,.
\end {equation}
The double lines represent the resummed gauge propagator (\ref{eq:G}).

   We have now discussed the cost of adding purely gauge loops to a diagram.
Before proceeding to the case of generic scalar loops in a diagram, it will
be useful to first review the leading-order calculation of
ref.~\cite{Arnold&Yaffe} for the
tricritical value of $\lambda/g^2$.


\subsection {Tricritical point at leading order}

   For a second-order phase transition, the transition occurs when the
effective mass of the scalar field vanishes.  A tricritical point occurs
when the effective, low-momentum, quartic coupling $\lambda\eff$
of the scalar vanishes
as well.  The classic mean-field example is the potential
\begin {equation}
   V = {\textstyle{1\over2}} m^2 |\vec\Phi|^2
       + {\textstyle{1\over4!}} \lambda |\vec\Phi|^4
       + {\textstyle{1\over6!}} \eta |\vec\Phi|^6 \,,
\end {equation}
which (ignoring corrections due to fluctuations) has a second-order
transition in $m^2$ if $\lambda > 0$, a first-order transition in $m^2$
if $\lambda < 0$, and a tricritical point at $\lambda = 0$.
The actual low-momentum effective potential for gauge-Higgs theories
was computed at leading order in $1/N$ in ref.~\cite{Arnold&Yaffe}.
Here, we just need the effective
value of the four-point interaction.
We begin by assuming $\lambda/g^2$ is $O(N^{-1})$,
which we shall see {\it a posteriori} is the correct place to look for
the tricritical point.  For simplicity, we will also ignore the
bare six-point coupling $\eta$ by setting it to zero.%
\footnote{
   This is sensible, based on (\protect\ref{eq:matching}), if one
   formally considers $g_4^2$ to be arbitrarily small.  If one
   instead considers the natural choice that $g_4^2$ is $O(N^{-1})$,
   then $\eta$ will be $O(N^{-3})$.  At the order under consideration,
   the effects of $\eta$ on the following derivation can be absorbed into
   the definition of $\lambda$.
}
The theory is then super-renormalizeable.  $\lambda$ and $g^2$ do not require
renormalization and will henceforth refer to their bare, short-distance
values.

   It is convenient to henceforth think of $g^2$ as
$O(N^{-1})$ and so $\lambda$ as $O(N^{-2})$.
This is just a convention because $g^2$ is dimensionful,
but it is a convenient convention because it makes the internal momenta
of gauge bosons in the graphs discussed above $O(N^0)$.
$N$ counting of those
graphs then reduces to counting scalar loops and explicit
coupling constants.

\begin {figure}
\vbox
   {%
   \begin {center}
      \leavevmode
      
      \epsfbox [150 370 500 450] {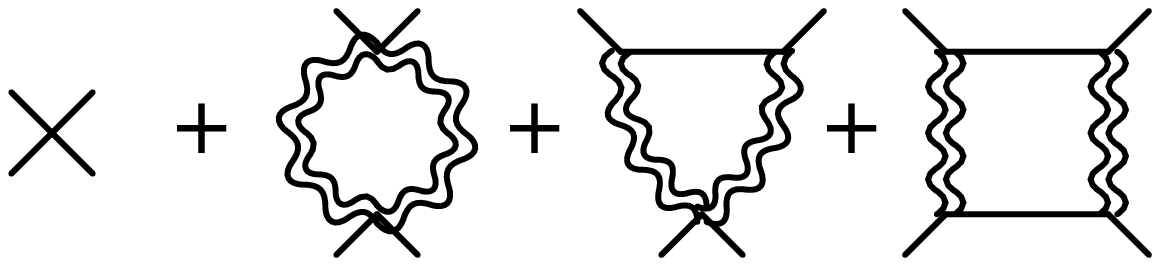}
   \end {center}
   \caption
       {%
       Leading-order contributions to the scalar four-point interaction.
       \label{fig:LO}
       }%
   }%
\end {figure}

The leading-order graphs for the four-point
interaction are shown in fig.~\ref{fig:LO},
where the double lines again represent
the resummed gauge propagator (\ref{eq:G}).  For the U(1) case, these
graphs give
\begin {equation}
   \lambda\eff = \lambda
        - 12 g^4 \int {d^3p\over(2\pi)^3} {1\over(p^2 + a N g^2 p)^2}
   = \lambda - {6 g^2\over\pi^2 a N} \,,
\end {equation}
where computation of the self-energy diagrams fig.~\ref{fig:pi} yields
\begin {equation}
   \Pi_{\mu\nu}(p) = a N g^2 p
       \left(\delta_{\mu\nu} - {p_\mu p_\nu \over p^2}\right) \,,
   \qquad\qquad
   a = {\textstyle{1\over16}} \,.
\end {equation}
Setting $\lambda\eff$ to zero, the tricritical point is at
\begin {equation}
   {\lambda\over g^2} = {96\over\pi^2 N} + O(N^{-2}) \,.
\end {equation}
The case of SU(2) with $N$ doublet Higgs bosons differs just by the number
of gauge bosons and the normalization of the coupling of the scalars to
the gauge bosons.  With conventional normalization of $g$,
\begin {equation}
   {\lambda\over g^2} = {36\over\pi^2 N} + O(N^{-2}) \,.
\end {equation}


\subsection {Scalar loops in the infrared}

  Consider for a moment pure scalar theory.  At zero external momentum,
the naive loop expansion parameter is $N \lambda/m$, where the $1/m$ follows
from dimensional analysis.  There is therefore an infrared problem when
one goes to the transition, $m\to0$.  The standard application of large
$N$ techniques to this theory eradicates this problem by a large $N$
resummation of the quartic interaction, which curbs the infrared
behavior of the diagrammatic expansion.%
\footnote{
   This is most easily achieved by the standard technique of replacing
   the quartic interaction $\lambda\phi^4$ by
   $\chi^2 + \sqrt\lambda\phi^2 \chi$
   where $\chi$ is an auxiliary field, integrating out $\phi$, and then
   studying the resulting theory of $\chi$.
}

  In gauge-Higgs theories, the same problem potentially arises and will
require a similar resummation of the scalar interactions.  However, for
the special case of computing the tricritical value of $\lambda/g^2$,
this resummation is unnecessary.  Because the problem with the naive
expansion in scalar loops was an infrared problem, the loop expansion
parameter $N\lambda/m$ due to the infrared behavior ($p \ll Ng^2$)
of loops should be replaced by $N\lambda\eff/m$.  But $\lambda\eff$
at the tricritical point is zero by definition.  As we have reviewed,
that zero occurs because of a cancelation of interactions, such as those
shown at leading order in fig.~\ref{fig:LO}.  This cancelation breaks down
for loop momenta $p \gtrsim Ng^2$, so the real cost of adding a scalar loop at
the tricritical point will be determined by the scale $N g^2$:
\begin {equation}
   {N \lambda \over p}  \sim  {\lambda\over g^2} \sim {1\over N} \,.
\end {equation}

\begin {figure}
\vbox
   {%
   \begin {center}
      \leavevmode
      
      \epsfbox [150 140 500 610] {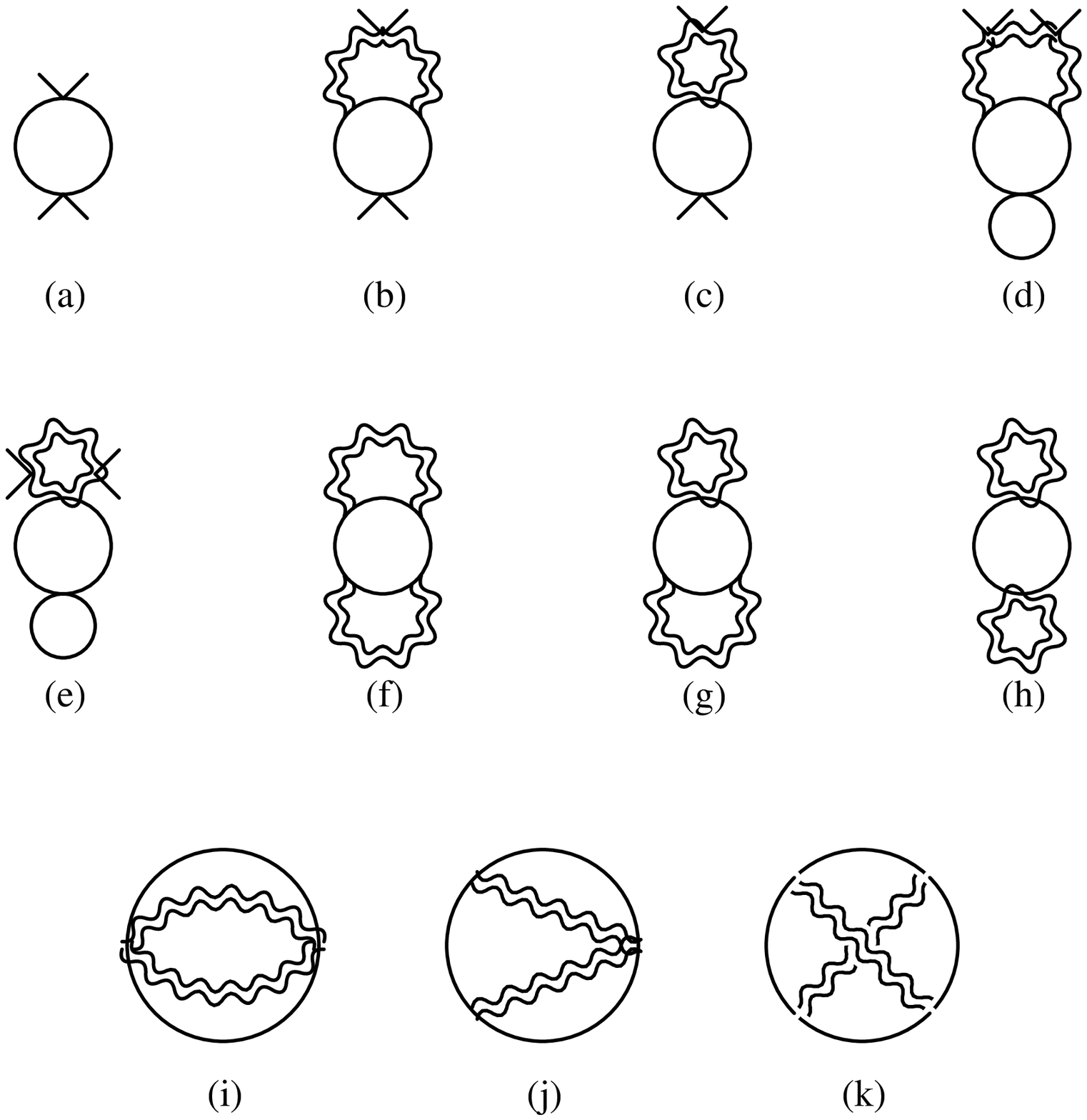}
   \end {center}
   \caption
       {%
       Next-to-leading order contributions to the scalar
       four-point interaction: abelian graphs.
       We have neglected graphs that vanish in the Abelian case
       due to Furry's theorem, {\it i.e.} charge conjugation.
       See fig.~\protect\ref{fig:NLO key} for interpretation of (f--k).
       \label{fig:NLO}
       }%
   }%
\end {figure}

\begin {figure}
\vbox
   {%
   \begin {center}
      \leavevmode
      
      \epsfbox [150 230 500 520] {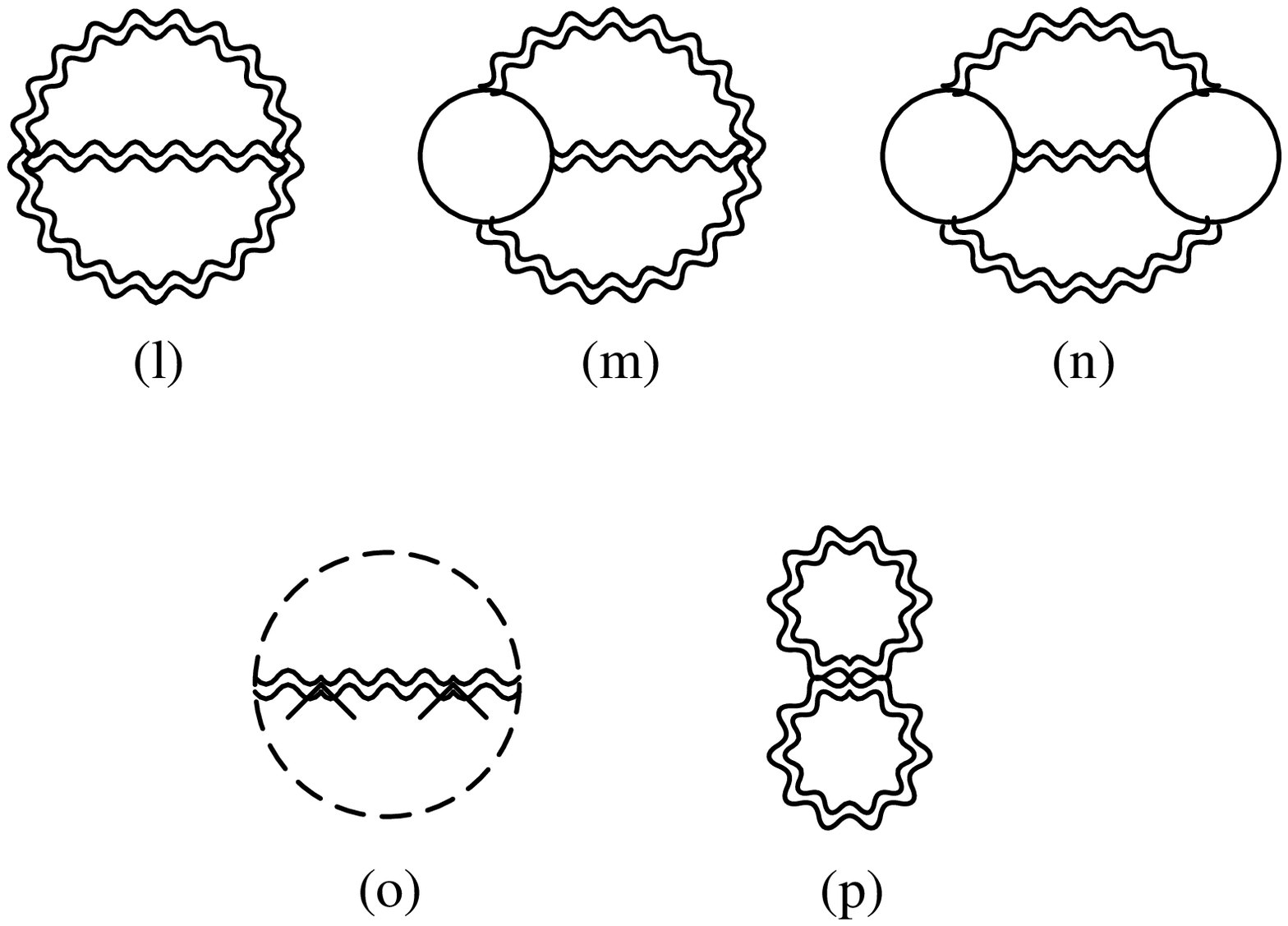}
   \end {center}
   \caption
       {%
       Next-to-leading order contributions to the scalar
       four-point interaction: SU(2) graphs in Landau gauge.
       We have neglected graphs which vanish by charge conjugation.
       Note that the subgraph of fig.~\protect\ref{fig:vanish}
       vanishes for SU(2) but not
       other groups.
       The dashed lines represent ghosts.
       See fig.~\protect\ref{fig:NLO key} for interpretation of (l--n,p).
       \label{fig:NLO2}
       }%
   }%
\end {figure}

\begin {figure}
\vbox
   {%
   \begin {center}
      \leavevmode
      
      \epsfbox [150 325 500 480] {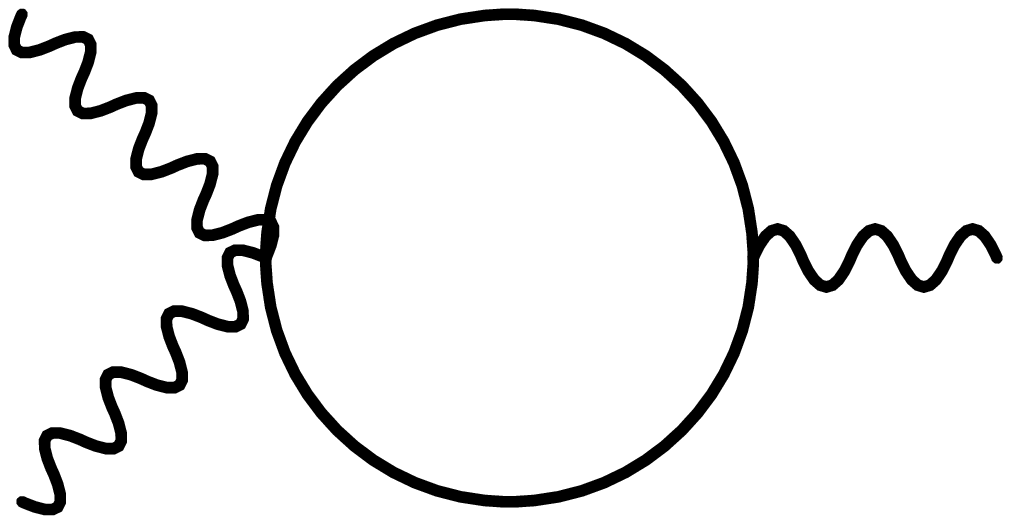}
   \end {center}
   \caption
       {%
       A subgraph which vanishes for U(1) and SU(2).
       \label{fig:vanish}
       }%
   }%
\end {figure}

\begin {figure}
\vbox
   {%
   \begin {center}
      \leavevmode
      
      \epsfbox [150 400 500 620] {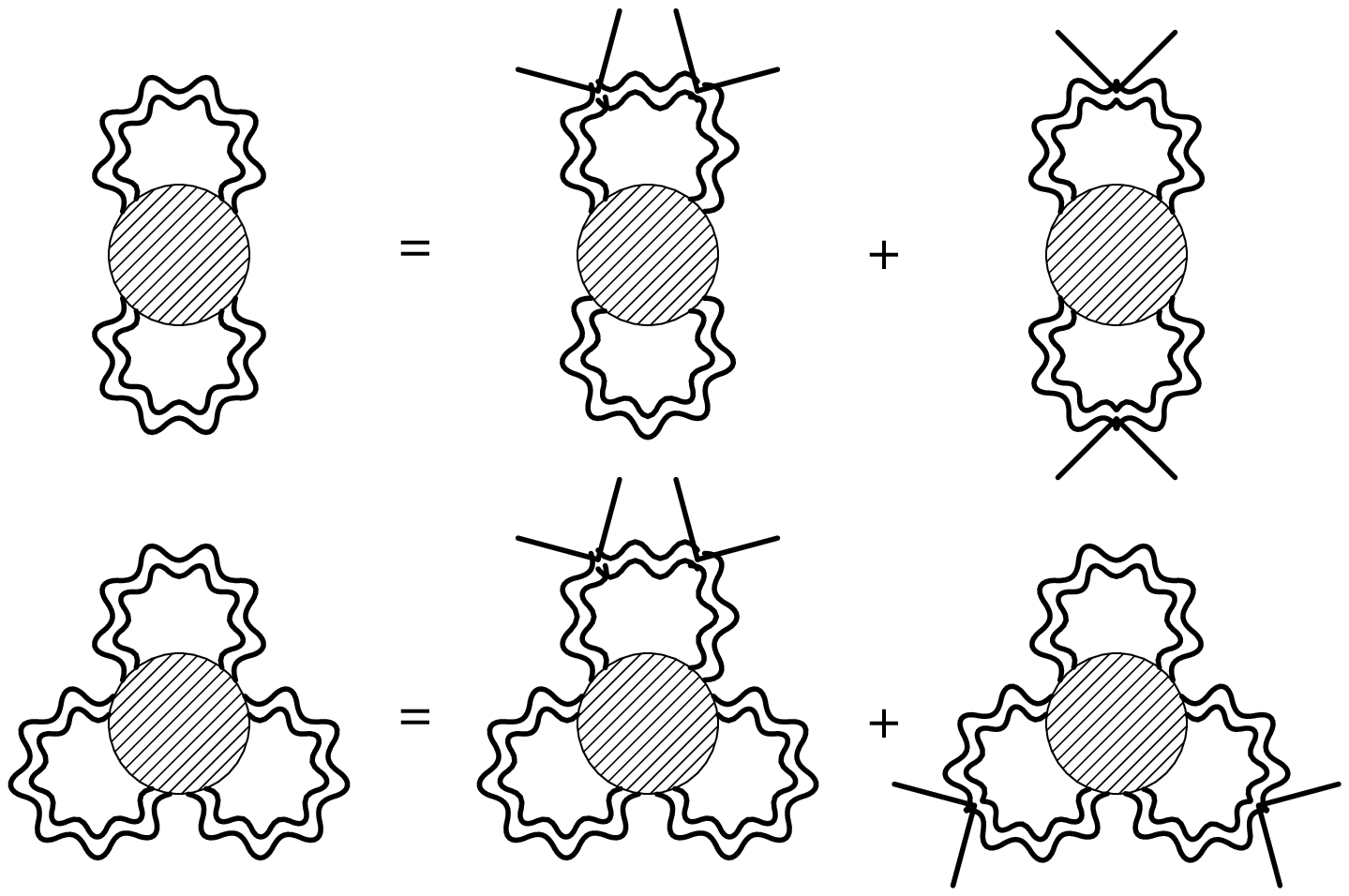}
   \end {center}
   \caption
       {%
       Meaning of abbreviated graphs in figs.~\protect\ref{fig:NLO}
       and \protect\ref{fig:NLO2}.
       \label{fig:NLO key}
       }%
   }%
\end {figure}

\begin {figure}
\vbox
   {%
   \begin {center}
      \leavevmode
      
      \epsfbox [150 380 500 420] {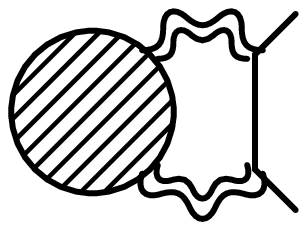}
   \end {center}
   \caption
       {%
       Example of diagrams which vanish in Landau gauge for zero external
       momentum.
       \label{fig:landau great}
       }%
   }%
\end {figure}

   The upshot is that, to determine the tricritical point, we can set $m{=}0$
and then use naive large $N$ power-counting of diagrams, treating
$g^2 \sim O(N^{-1})$ and $\lambda \sim O(N^{-2})$.  The relevant
diagrams for computing $\lambda\eff$ at next-to-leading order in
$1/N$ are shown in figs.~\ref{fig:NLO} and \ref{fig:NLO2} for
Landau gauge.
In this gauge, diagrams of the form of fig.~\ref{fig:landau great} vanish
at zero external momentum.
The graphs (f--n,p) denote
the contributions to $\lambda\eff$ in the short-hand style explained by
fig.~\ref{fig:NLO key}.  At each order in the calculation of
$\lambda\eff$, and the simultaneous determination of the tricritical point by
setting $\lambda\eff=0$, one will find that scalar infrared divergences
always cancel at the order one is calculating.  We now turn to a
calculation of the U(1) case, corresponding to the diagrams
fig.~\ref{fig:NLO}, where this will be made explicit.


\section {U(1) tricritical point: next-to-leading order}

\begin {figure}
\vbox
   {%
   \begin {center}
      \leavevmode
      
      \epsfbox [150 370 500 440] {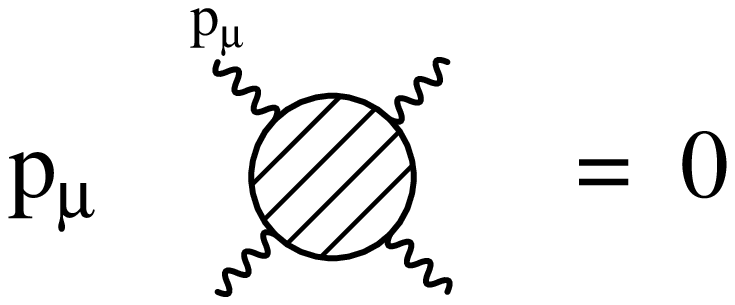}
   \end {center}
   \caption
       {%
       A U(1) Ward identity.
       \label{fig:Ward}
       }%
   }%
\end {figure}

   For simplicity, we shall present the calculation in Feynman gauge.
The total contribution of diagrams of the form of fig.~\ref{fig:landau great}
still vanishes
in the U(1) case because of the U(1) Ward identity of fig.~\ref{fig:Ward}.%
\footnote{
   We have checked this explicitly.  We have also explicitly checked our
   final U(1) results are the same in any covariant gauge.
}
(Landau gauge results, which will be useful for the SU(2) case,
are given in Appendix~\ref{apndx:Landau}.)

   We shall proceed by doing the scalar loop integrals of all the diagrams
of fig.~\ref{fig:NLO}, summing up the diagrams, and then integrating over
the gauge boson momenta.  The order of the last two steps is important
because the gauge momentum integrals are infrared divergent for the
individual diagrams but not for the sum, and it will be convenient not
to have to introduce a consistent infrared regulator.
(Our disregard of regularization will sometimes be a bit cavalier in
this section, and we delay discussion of potential subtleties to
section 4.)

   As a warmup, consider fig.~\ref{fig:NLO}i.
This diagram gives a contribution to $\lambda\eff$ of
\begin {equation}
   \dlam^{\rm (i)} = - 72 N g^8 \intFF \int_l
         {1\over l^2(\v l + \v p + \v q)^2} \,
   = - {N g^8} \intFF {9\over|\v p + \v q|} \,,
\label{eq:Vg}
\end {equation}
where we have introduced the notation
\begin {eqnarray}
   \int_p &\equiv& \int {d^3p\over(2\pi)^3} \,,
\\
   {\cal F}_{pq} &\equiv& f_p^3 f_q + f_p^2 f_q^2 + f_p f_q^3 \,,
\end {eqnarray}
and $f_p$ is the large $N$ resummed gauge propagator
\begin {equation}
   f_p \equiv {1 \over p^2 + a N g^2 p} \,.
\end {equation}
A useful table of various $l$ integrals is given in
Appendix~\ref{apndx:integrals}.

   As a slightly more complicated example, consider fig.~\ref{fig:NLO}j,
which gives
\begin {equation}
   \dlam^{\rm (j)} = 48 N g^8 \intFF \int_l
       { (2\v l-\v p)\cdot(2\v l+\v q)
         \over l^2 (\v l - \v p)^2 (\v l + \v q)^2
       } \,.
\label{eq:Vh0}
\end {equation}
To simplify the $l$ integration, one may use the standard technique of
rewriting numerators in terms of denominators and things that don't
involve $l$:
\begin {equation}
   (2\v l-\v p)\cdot(2\v l+\v q)
       = (\v l-\v p)^2 + (\v l + \v q)^2 + 2 l^2
           - (p^2 + \v p\cdot\v q + q^2) \,,
\end {equation}
giving
\begin {eqnarray}
   \dlam^{\rm (j)} &=& {48 N g^8} \intFF \int_l \left[
         {1\over l^2(\v l + \v q)^2}
       + {1\over l^2(\v l - \v p)^2}
       + {2\over (\v l - \v p)^2 (\v l + \v q)^2}
       - {(p^2 + \v p\cdot\v q + q^2)\over
                 l^2 (\v l - \v p)^2 (\v l + \v q)^2}
     \right]
\nonumber\\
   &=& {N g^8} \intFF \left[
         {6\over q} + {6\over p}
       + {12\over|\v p + \v q|}
       - {6(p^2 + \v p\cdot \v q + q^2)\over pq|\v p + \v q|}
     \right]
\label{eq:Vh}
\end {eqnarray}
Appendix~\ref{apndx:integrals} explains an amusingly simple method for
evaluating the last $l$ integral by using a simple change of variables.

   Fig.~\ref{fig:NLO}k can be done similarly, and the result is given in
Appendix~\ref{apndx:summary}.

   Figs.~\ref{fig:NLO}(a--h) are slightly more subtle
because the scalar integration
is infrared divergent, both diagram by diagram and collectively.  For instance
fig.~\ref{fig:NLO}a gives
\begin {equation}
   \dlam^{\rm (a)} = -{\lambda^2\over3} \int_l {1\over l^4} \,.
\end {equation}
However, as we discussed earlier, $\lambda$ should end up replaced by
$\lambda\eff=0$ in the infrared if we sum up diagrams, as was shown at
leading-order in fig.~\ref{fig:LO}.
The cancelation in fig.~\ref{fig:LO} will correspond
to a cancelation, at this order in $1/N$, in the diagrams of
fig.~\ref{fig:NLO}(a--c,f--h)
if we consider the pieces of (f--h) represented by the
{\it second} term on the right-hand side of fig.~\ref{fig:NLO key}a.%
\footnote{
   Keep in mind that the last two terms of fig.~\ref{fig:LO} can be ignored
   on the context of the NLO diagrams (a--h) because of the Ward identity
   of fig.~\ref{fig:Ward}.
}
That is, if we set
\begin {equation}
   \lambda = {96\over\pi^2 N} g^2 + O(N^{-2})
\end {equation}
to make $\lambda\eff$ zero at leading order, then we will find that
the infrared divergences just discussed will cancel each other at the
order in $1/N$ at which we are computing.  It will be convenient to
write this condition at a more primitive level,
related directly to the diagrams of fig.~\ref{fig:LO}, as
\begin {equation}
   \lambda = 12 C_\lambda g^4 \int_p f_p^2 \,,
\end {equation}
where the value of $\lambda$ is now parametrized by $C_\lambda$ and
\begin {equation}
   C_\lambda = 1 + O(N^{-1})
\end {equation}
at the tricritical point.

\begin {figure}
\vbox
   {%
   \begin {center}
      \leavevmode
      
      \epsfbox [150 360 500 430] {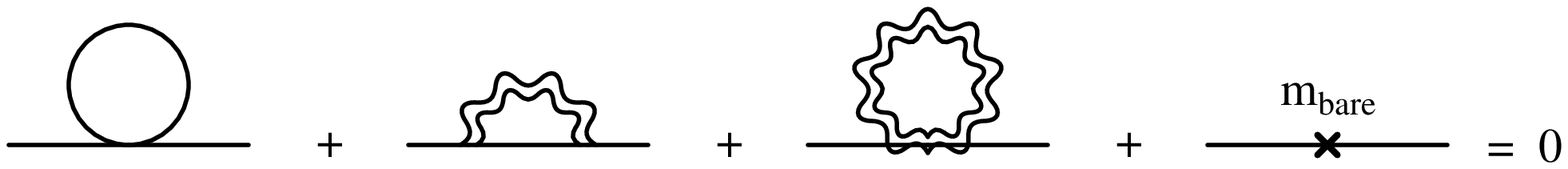}
   \end {center}
   \caption
       {%
       Diagrams contributing to the effective scalar mass $m\eff$.
       \label{fig:mass}
       }%
   }%
\end {figure}

\begin {figure}
\vbox
   {%
   \begin {center}
      \leavevmode
      
      \epsfbox [150 300 500 470] {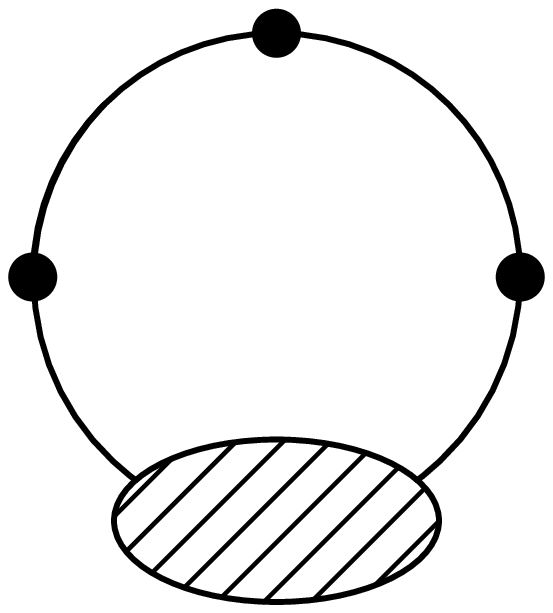}
   \end {center}
   \caption
       {%
       IR divergent loops caused by treating a non-zero mass
       perturbatively.  The circles represent mass insertions on a
       massless propagator.
       \label{fig:mass trouble}
       }%
   }%
\end {figure}

\begin {figure}
\vbox
   {%
   \begin {center}
      \leavevmode
      
      \epsfbox [120 390 470 500] {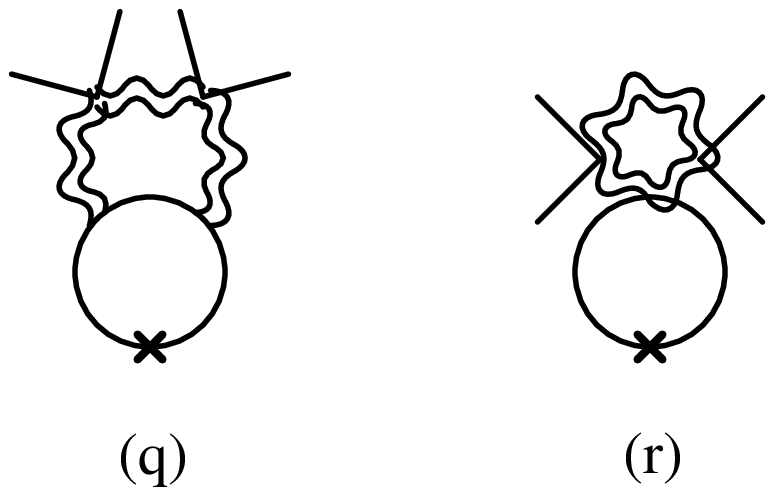}
   \end {center}
   \caption
       {%
       Additional NLO diagrams, containing bare mass insertion.
       \label{fig:counterterm}
       }%
   }%
\end {figure}

   There still remain infrared divergences, however, in the pieces of
figs.~\ref{fig:NLO}(d--h) corresponding to the {\it first} term on the
right-hand side of fig.~\ref{fig:NLO key}a.
The divergence does not cancel in the sum of these graphs because,
though we have taken $m=0$ in our perturbative scalar propagators, we have
so far ignored the fact that radiative corrections such as the first
three diagrams of fig.~\ref{fig:mass} will generate a contribution to the mass.
And if the mass of a massive scalar is treated perturbatively, it will
generate infrared divergences,
such as depicted by fig.~\ref{fig:mass trouble}, which shows
massless propagators connecting a perturbative mass insertion.
To be at the transition $m\eff=0$, we need to fine-tune the bare mass
at this order to satisfy the equation of fig.~\ref{fig:mass}.
An application of the Feynman rules
for these diagrams shows that this requires
\begin {equation}
   m\bare = -2 g^2 \int_p f_p
     ~~-~~ {\textstyle{2\over3}}N\lambda \int_p {1\over p^2}
     ~~+~~ \hbox{(higher order)}\,.
\label {eq:mbare}
\end {equation}
If we treat $m\bare$ perturbatively, it will cancel the radiative
contributions to the mass order by order in perturbation theory.
All we need to do to cancel the infrared divergences in our calculation
of $\lambda\eff$ at this order is to include the additional diagrams of
fig.~\ref{fig:counterterm}.  We will henceforth ignore the graphs
of figs.~\ref{fig:NLO}d and e and the second term of the bare mass
(\ref{eq:mbare}) above, as these trivially cancel each other.

   The graph of fig.~\ref{fig:NLO}f gives a contribution to $\dlam$ of
\begin {mathletters}%
\label{eq:div}
\begin {equation}
   -12 N g^8 \intFF \int_l
       { (2\v l - \v p)^2 (2\v l + \v q)^2
            \over l^4 (\v l - \v p)^2 (\v l + \v q)^2 } \,;
\label{eq:div1}
\end {equation}
the graphs of figs.~\ref{fig:NLO}(b,g) and
\ref{fig:counterterm}\countera\ give
\begin {equation}
   -12 N g^8 \int_{pq}
       \left[ f_p f_q^3 + (3-2C_\lambda) f_p^2 f_q^2 + 3 f_p^3 f_q \right]
       \int_l \left[ - {(2\v l+\v q)^2\over l^4 (\v l + \v q)^2} \right]
   ~~-12 N g^8 \{ p \leftrightarrow q \}
  \,;
\label{eq:div2}
\end {equation}
and the graphs of figs.~\ref{fig:NLO}(a,c,h) and
\ref{fig:counterterm}\counterb\  give
\begin {equation}
   -12 N g^8 \int_{pq}
       \left[ 3 f_p f_q^3 + (3-2C_\lambda)^2 f_p^2 f_q^2 + 3 f_p^3 f_q \right]
       \int_l {1\over l^4} \,.
\label{eq:div3}
\end {equation}%
\end{mathletters}%
Now consider the sum of (\ref{eq:div}a--c).
Setting $C_\lambda$ to $1$, we find that the $l$ integral of the sum of
the integrands converges, as promised.%
\footnote{
   More accurately, it converges if one ignores logarithmic infrared
   divergences
   $\int_l (\v l/l^4)$ that vanish by parity.  The physical regulator---an
   arbitrarily tiny mass term due to being infinitesimally above the
   transition---respects parity.  Dimensional regularization does also.
}

   Having verified that the infrared divergences cancel, it is convenient
to proceed by computing the various terms individually, regulating the
$l$ integration with dimensional regularization.  The results are given
in Appendix~\ref{apndx:summary}.

   The next step is to do the $p$ and $q$ integrations.  We do this by
first performing the integration over the relative angle $\theta$
between $\v p$
and $\v q$.  As an example, consider fig.~\ref{fig:NLO}j again and our result
(\ref{eq:Vh}).  By using the angular averages
\begin {equation}
   \left\langle {1\over|\v p + \v q|} \right\rangle_\theta = {1\over p_>}
   \qquad \hbox{and} \qquad
   \left\langle {\cos\theta \over|\v p + \v q|} \right\rangle_\theta
            = - {p_< \over 3 p_>^2} \,,
\end {equation}
we obtain
\begin {equation}
   \dlam^{\rm(j)} = N g^8 \intFF {1\over p_>} \left(
      18 - 4x
   \right) \,,
\label{eq:Vh2}
\end {equation}
where
\begin {equation}
   p_> \equiv \max(p,q) \,,
   \qquad
   p_< \equiv \min(p,q) \,,
   \qquad
   x \equiv p_< / p_> \,.
\end {equation}
Similar results for the rest of the graphs are given in
Appendix~\ref{apndx:summary}.
The sum of all graphs gives
\begin {eqnarray}
\dlam &=& N g^8 \intFF {h(x)\over p_>}
   + N g^8 \int_{pq} f_p^3 f_q \left[ {12\over q} \right]
   \,,
\label{eq:dlam}
\\
h(x) &=&
   {\textstyle -6 + {11\over2}x + 3x^{-1}
    - \left({15\over2} + 3x^{-2} + 3x^2\right)}
         (1+x^2)^{-1/2} \, {\rm Sinh}^{-1}x
   \,.
\label {eq:h}
\end {eqnarray}
To do one of the integrals easily, write
\begin {eqnarray}
   \intFF {h(x)\over p_>}
   &=& {1\over2\pi^4} \int_0^\infty p^2 dp
       \int_0^p q^2 dq \, \F_{pq} \, {h(q/p)\over p}
\nonumber\\
   &=& {1\over 2\pi^4} \int_0^1 dx x^2 h(x)
        \int_0^\infty dp \, p^4 (f_p f_{xp}^3 + f_p^2 f_{xp}^2 + f_p^3 f_{xp})
\nonumber\\
   &=& {-1\over 2\pi^4 a^3 N^3 g^6} \int_0^1 {dx\over x}
         \left[{(1+x)\over(1-x)} \ln x + {3\over2}\right]
         (1+x) h(x)
   \,,
\end {eqnarray}
and similarly
\begin {equation}
   \int_{pq} f_p^3 f_q {1\over q}
   = {-1\over4\pi^4 a^3 N^3 g^6} \int_0^1 {dx\over x (1-x)^2}
         \left[{(1+x^3)\over(1-x)} \ln x
              + {3\over2} - x + {3\over2} x^2 \right]
   .
\end {equation}
Combining them gives
\begin {equation}
  \dlam = -\, {\kappa g^2 \over 2\pi^4 a^3 N^2}
               = -\, {2048\, \kappa g^2\over \pi^4 N^2} \,,
\end {equation}
where
\begin {eqnarray}
   \kappa &=& \int_0^1 {dx\over x} \biggl\{
      \left[{(1+x)\over(1-x)}\ln x + {3\over2}\right] (1+x) h(x)
\nonumber\\ && \qquad\qquad
   + {6\over(1-x)^2} \left[ {(1+x^3)\over(1-x)} \ln x 
      + {3\over2} - x + {3\over2} x^2 \right] \biggr\} \,.
\end {eqnarray}
The contributions of each diagram to $\kappa$ contain logarithmic
divergences at small $x$, but the total is integrable.
The integral can in principal be done analytically, with the
result expressed in terms of generalized polylogarithms with
arguments like $\sqrt2$, but this seems unhelpful enough that we haven't
bothered.%
\footnote{
   The first step is to change variables to $y = x + \sqrt{1+x^2}$ in
   the terms involving ${\rm Sinh}^{-1}x$, which then becomes $\ln y$.  This
   transforms the integral into a product of rational functions and logarithms.
}
Numerical integration gives
\begin {equation}
   \kappa = 1.68536...
\end {equation}
The final results for the tricritical value of $\lambda/g^2$ is
then
\begin {eqnarray}
   {\lambda/g^2}
   &=& {96\over\pi^2} \left[
      N^{-1} + {64\kappa\over3\pi^2} N^{-2} + O(N^{-3}) \right]
\nonumber\\
   &=& {96\over\pi^2} \left[
      N^{-1} + 3.64293 N^{-2} + O(N^{-3}) \right] .
\end{eqnarray}


\section {Some regularization issues}

\begin {figure}
\vbox
   {%
   \begin {center}
      \leavevmode
      
      \epsfbox [150 300 500 500] {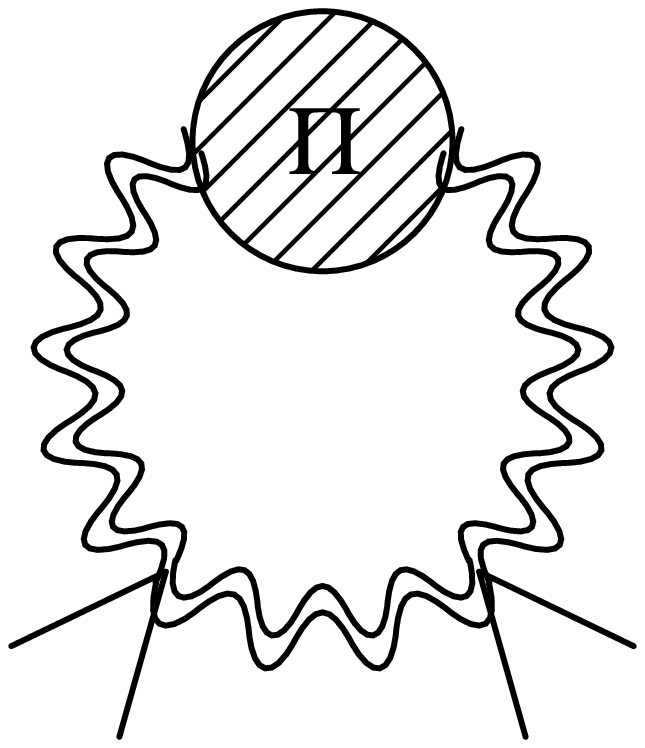}
   \end {center}
   \caption
       {%
       A class of diagram with IR divergences in the gauge momentum.
       \label{fig:divergences}
       }%
   }%
\end {figure}

   In the last section, we did not bother to introduce a consistent
regularization of divergences diagram by diagram, arguing that
that divergences cancel when all diagrams were summed.  This is a
potentially dangerous argument and shall later plague us in the
SU(2) calculation if we do not address it.  To see the problem,
consider the contributions of the form of fig.~\ref{fig:divergences}.
The gauge
loop integration would have an infrared divergence of the form
\begin {equation}
   \int_p f_p^3 \Pi(0)
\label{eq:Pi0 divergence}
\end {equation}
if the self-energy $\Pi$ did not vanish at zero momentum.
Fortunately, $\Pi(0) = 0$ is a consequence of Ward identities.
The subtlety arises because $\Pi(0)$ does not necessarily vanish diagram by
diagram.  As an example, consider the
{\it one}-loop contribution to $\Pi(0)$, as shown in
fig.~\ref{fig:pi}.
Using dimensional regularization for the UV and a small scalar
mass for the IR, these diagrams give
\begin {equation}
   \Pi_{\mu\mu}(0) \approx -4 g^2 \int_l {l^2 \over (l^2+m^2)^2}
                           +2 d g^2 \int_l {1\over l^2} \,.
\label{eq:IR example}
\end {equation}
If we ignore the UV regularization and instead set $d=3$ and combine the
integrands, our $l$ integral is {\it still} UV divergent.  Our
calculation (\ref{eq:Pi0 divergence}) would then end up with
ill-defined products of IR and UV divergences of the form
\begin {equation}
   \int_{\rm IR} {d^3p \over p^3} \int_{\rm UV} {d^3l \over l^2} \,.
\end {equation}

The solution to avoiding this problem is to return to consistently
regulating the theory.  Then rewrite fig.~\ref{fig:divergences} as
\begin {equation}
   \int_p f_p^3 \Pi(0) = \int_p f_p^3 [\Pi(p) - \Pi(0)]
            + \int_p f_p^2 \Pi(0) \,.
\end {equation}
The $p$ integration for the first term on the right is now IR convergent,
and the calculation of $\Pi(p)-\Pi(0)$ is UV convergent diagram by
diagram; so we can set $d=3$ in this calculation and proceed as we did
in the previous section.  The second term on the right-hand side requires
full regularization, but we know from Ward identities that $\Pi(0)=0$,
so we do not need to calculate it.  The final prescription, then, is
to replace $\Pi(p)$ by $\Pi(p)-\Pi(0)$ when computing diagrams like
fig.~\ref{fig:divergences}.

The effect of this prescription on the U(1) calculation is that we
should add by hand to (\ref{eq:dlam}) a term proportional to
\begin {equation}
   \int_{pq} f_p^3 f_q \left[1\over q\right]
\end {equation}
in order to remove any remaining IR/UV divergences in the $p$/$q$ integrals.
However, no such term is needed, and so the prescription has no effect
on our previous calculation.%
\footnote{
   This isn't an accident.  The two loop contribution to $\Pi_{\mu\nu}(0)$ are
   UV log divergent diagram by diagram.  For log divergences, cancelations
   are maintained even when the UV regularization is removed.  Consider the
   example of (\ref{eq:IR example}) for $d{=}2$.
}
As we shall see, however, the prescription will be important to
the SU(2) calculation of the next section.


\section {SU(2) tricritical point: next-to-leading order}

\begin {figure}
\vbox
   {%
   \begin {center}
      \leavevmode
      
      \epsfbox [150 320 500 500] {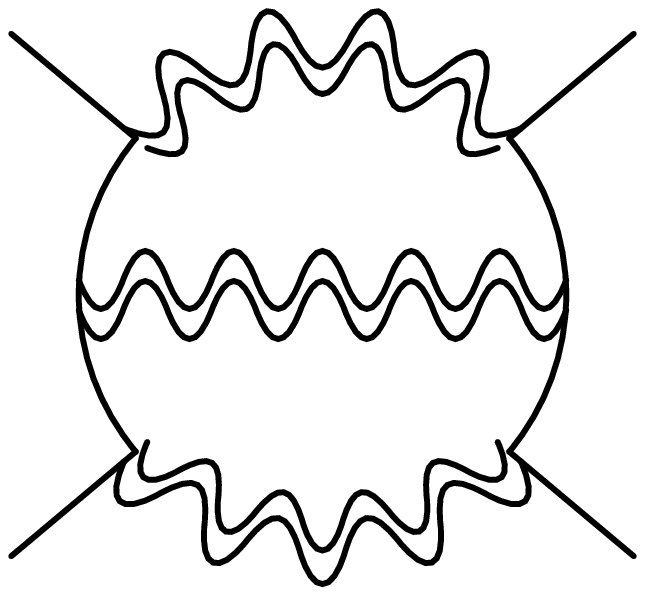}
   \end {center}
   \caption
       {%
       A diagram which vanishes in Landau gauge.
       \label{fig:hooray Landau}
       }%
   }%
\end {figure}

   The SU(2) case is more convenient to treat in Landau gauge than in
Feynman gauge.  This is because there are a host of diagrams, such as
fig.~\ref{fig:hooray Landau},
which vanish in Landau gauge for zero external momentum
(because the gluon polarization the external scalars couple to is
proportional to the gluon four-momentum and does not propagate in Landau
gauge).

\begin {figure}
\vbox
   {%
   \begin {center}
      \leavevmode
      
      \epsfbox [150 360 500 420] {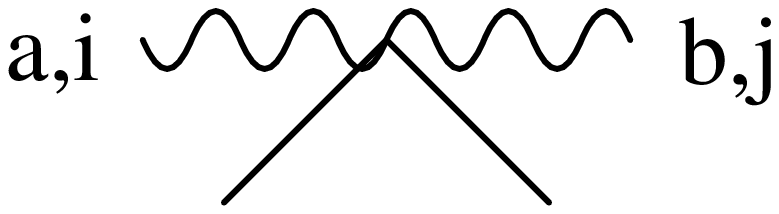}
   \end {center}
   \caption
       {%
       Scalar insertion on gauge line.
       \label{fig:scalar insertion}
       }%
   }%
\end {figure}

   We will restrict our attention to the group SU(2) because it has
the convenient property that scalar insertions such as shown in
fig.~\ref{fig:scalar insertion}
are proportional to $\delta_{ab}\delta_{ij}$.  This greatly simplifies the
analysis of the group factors in diagrams.%
\footnote{
   The quartic interaction $|\v\Phi|^4$ in (\protect\ref{eq:S3}) will mean
   $(\v\Phi^\dagger_\alpha\cdot\v\Phi^\alpha)^2$,
   where $\alpha$ is the SU(2) index.
   As noted in ref.~\protect\cite{Arnold&Yaffe},
   the theory with this interaction has a bigger global symmetry
   than simple flavor
   U($N$).
   The scalar sector has an O($2N$) symmetry.  The gauge interactions
   reduce the symmetry to ${\rm SU}(2)_{\rm L} \times {\rm Sp}(2N)_{\rm R}$,
   where ${\rm Sp}(2N)_{\rm R}$ is the $N$-flavor generalization
   of the usual ${\rm SU}(2)_{\rm R} \simeq {\rm Sp}(2)_{\rm R}$
   custodial symmetry of the $N{=}1$ case.  This symmetry is larger
   than U($N$) and requires that
   the scalar potential be a function of only the single variable
   $\v\Phi^\dagger_\alpha\cdot\v\Phi^\alpha$, forbidding other possibilities
   such as
   $(\v\Phi^\dagger_\alpha\cdot\v\Phi^\beta)
    (\v\Phi^\dagger_\beta\cdot\v\Phi^\alpha)$.
   The appearance of such couplings severely complicates the
   generalization of our treatment to general gauge groups.
}

   The Landau gauge results for the Abelian diagrams of
fig.~\ref{fig:NLO}
are given in Appendix~\ref{apndx:Landau}.
For SU(2), they give the contribution
\begin {equation}
\dlam^{\rm(a-k,q,r)} = N g^8 \intFF {h_1(x)\over p_>}
   + N g^8 \int_{pq} f_p^3 f_q \left[ {27\over32 q} \right]
   \,,
\end{equation}
\begin {equation}
h_1(x) =
   {\textstyle -{27\over64} + {15\over256}x - {9\over160}x^2
    - {9\over128}x^{-1}
    + \left({45\over256} + {9\over128}x^{-2} + {9\over128}x^2\right)}
         (1+x^2)^{-1/2} \, {\rm Sinh}^{-1}x
   \,.
\label {eq:h1}
\end {equation}
In the SU(2) case,
\begin {equation}
   a = {1\over32} \,,
\end {equation}
and then
\begin {equation}
  \dlam^{\rm(a-k)} = -\, {\kappa_1 g^2 \over 2\pi^4 a^3 N^2}
               = -\, {2^{14}\, \kappa_1 g^2\over \pi^4 N^2} \,,
\end {equation}
where
\begin {eqnarray}
   \kappa_1 &=& \int_0^1 {dx\over x} \biggl\{
      \left[{(1+x)\over(1-x)}\ln x + {3\over2}\right] (1+x) h_1(x)
\nonumber\\ && \qquad\qquad
   + {27\over64(1-x)^2} \left[ {(1+x^3)\over(1-x)} \ln x 
      + {3\over2} - x + {3\over2} x^2 \right] \biggr\} \,.
\nonumber\\
   &=& -0.052539(1)
\end {eqnarray}

Now consider the non-Abelian graphs, and start with the ghost graph of
fig.~\ref{fig:NLO2}o.  Doing the ghost loop integration, one finds
\begin {equation}
   \dlam^{\rm(o)} = f^{abc} f^{abc} g^2 \left[
             - {3 \over512\pi^2 a^2 N^2}
             - {3\over 8} \int_p f_p^3 \int_q {1\over q^2}
           \right]
\label{eq:dlam m}
\end {equation}
By our prescription for handling divergences, discussed in the previous
section, the second term on the right-hand side should be discarded.
For SU(2),
\begin {equation}
   f^{abc} f^{abc} = 6 \,.
\end {equation}

The figure-eight graph of fig.~\ref{fig:NLO2}p gives
\begin {equation}
   \dlam^{\rm(p)} = f^{abc} f^{abc} g^2 \left[
             {1\over4\pi^4 a^2 N^2}
             + \int_{pq} f_q^3 f_p \left(
                 {\textstyle{9\over4}}
                 - {\textstyle{3\over4}} \cos^2\theta
             \right)
           \right]
\label{eq:dlam n}
\end {equation}
Again, the second term is thrown away by our prescription.

Finally, we have the graphs of fig.~\ref{fig:NLO2}(l-n).
Fig.~\ref{fig:NLO2}(n), for example, is
\begin {eqnarray}
   \dlam^{\rm(n)} &=& -{1\over32} f^{abc} f^{abc} g^6 \phi^4
       \int_{pq} \F_{pqr}
\nonumber\\ &&
    \times \int_{l_1}
       {(2l_1-p)_\mu (2l_1+q)_\nu (2l_1-p+q)_\rho
           \over l_1^2 (\v l_1-\v p)^2 (\v l_1+\v q)^2}
    \int_{l_2}
       {(2l_2-p)_{\bar\mu} (2l_2+q)_{\bar\nu}
        (2l_2-p+q)_{\bar\rho}
           \over l_2^2 (\v l_2-\v p)^2 (\v l_2+\v q)^2}
\nonumber\\ &&
    \times
    \left( \delta_{\mu\bar\mu} - {p_\mu p_{\bar\mu} \over p^2} \right)
    \left( \delta_{\nu\bar\nu} - {q_\nu q_{\bar\nu} \over q^2} \right)
    \left( \delta_{\rho\bar\rho}
                   - {r_\rho r_{\bar\rho} \over r^2} \right) \,,
\end {eqnarray}
where
\begin {equation}
   \v r = \v p + \v q
\end {equation}
and
\begin {equation}
   \F_{pqr} = 3 f_p^3 f_q f_r + 3 f_p^2 f_q^2 f_r \,.
\end {equation}
(If one prefers, one can symmetrize the definition of $\F_{pqr}$
with respect to permutations.)
We have evaluated the scalar integrals by brute force,
and various Feynman and momentum integrals required for this evaluation
are tabulated in appendices~\ref{apndx:Feyn param}
and \ref{apndx:bleah}.
The complicated results of the scalar integrations were contracted
with the Landau gauge vector propagators using a symbolic manipulation
program.
The result can be most simply expressed
when all three graphs are combined:
\begin {equation}
   \dlam^{\rm(l,m,n)} =
      f^{abc} f^{abc} g^6 \int_{pq} \F_{pqr} h_2(p,q,r) \,,
\label {eq:dlam jkl}
\end {equation}
where
\begin {eqnarray}
   h_2(p,q,r) &=& - (p+q-r) (p-q+r) (-p+q+r)
\nonumber\\ && \qquad
   \times \Biggl[
       {2 \mu^2 \over (p+q+r)^3}
       + {\mu (p+q+r+2\mu) \over 4pqr}
\nonumber\\ && \qquad\qquad
      + {(p+q+r+2\mu)^2\over 32p^2q^2r^2(p+q+r)}
           \left(p^4+q^4+r^4+6p^2q^2+6p^2r^2+6q^2r^2\right)
   \Biggr]
\end {eqnarray}
and
\begin {equation}
   \mu \equiv a N g^2 \,.
\end {equation}

Our prescription for dealing with divergences replaces
\begin {eqnarray}
   \dlam^{\rm(l,m,n)} &\to&
      f^{abc} f^{abc} g^6 \int_{pq} \left[
         \F_{pqr} h_2(p,q,r) + 3 f_p^3 q^{-2} \sin^2\theta_{pq} \right]
\nonumber\\
   &=& f^{abc} f^{abc} g^6 \mu^{-2} \kappa_2 \,,
\end {eqnarray}
where
\begin {eqnarray}
   \kappa_2 &=& {1\over8\pi^4} \int_0^\infty dp \, dq \int_{-1}^1 d(\cos\theta)
         \left[ p^2 q^2 \F_{pqr} h_2(p,q,r) + 3 p^2 f_p^3 \sin^2\theta
         \right]_{\mu{=}1}
\nonumber\\
   &=& -0.01573(1)
\end {eqnarray}
and the value of $\kappa_2$ has been obtained by direct numerical
integration.%
\footnote{
   The integral can be reduced to a two-dimensional integral by
   the rewriting $p_< = x p_>$ and integrating analytically over $p$,
   giving something too ugly to reproduce here.
}

Putting everything together, the final result for the tricritical value of
$\lambda/g^2$ is
\begin {eqnarray}
   {\lambda/g^2}
   &=& {36\over\pi^2} \left[
      N^{-1}
      - \left(
          {512\over3}\pi^2\kappa_2
          - {4096\over9\pi^2}\kappa_1
          - 1
          + {128\over3\pi^2}
      \right) N^{-2}
      + O(N^{-3}) \right]
\nonumber\\
   &=& {36\over\pi^2} \left[
      N^{-1} + 20.8\, N^{-2} + O(N^{-3}) \right] .
\end{eqnarray}



\bigskip

We thank Larry Yaffe, Lowell Brown, David Boulware, and Krishna
Rajagopal for useful conversations.
This work was supported by the U.S. Department of Energy,
grant DE-FG03-96ER40956.


\newpage
\appendix

\section {Useful integrals}
\label {apndx:integrals}

\subsection {Scalar integrals}
\label{apdnx:scalar integrals}

\begin {equation}
   z_{ij} \equiv |\v p_i-\v p_j|
\end {equation}
\begin {eqnarray}
 \int_l {1\over(\v l-\v p_1)^2(\v l-\v p_2)^2} &&=
    {1\over8 z_{12}}
\label{eq:bubble}
\\
 \int_l {1\over(\v l-\v p_1)^2(\v l-\v p_2)^2(\v l-\v p_3)^2} &&=
    {1\over8 z_{12}z_{23}z_{31}}
\label{eq:triangle}
\\
 \int_l {1\over l^2(\v l-\v p_1)^2(\v l-\v p_2)^2
               (\v l-\v p_1-\v p_2)^2} &&=
    {1\over8 p_1 p_2 \v p_1\cdot\v p_2}
    \left( {1\over|\v p_1-\v p_2|} - {1\over|\v p_1 + \v p_2|} \right)
\label{eq:x integral}
\\
\noalign{\hbox{Using dimensional regularization for the infrared:}}
 \int_l {1\over l^4} &&= 0
\\
 \int_l {1\over l^4(\v l-\v p)^2} &&= 0
\\
 \int_l {1\over l^4(\v l-\v p_1)^2(\v l-\v p_2)^2} &&=
    {\v p_1\cdot\v p_2 \over 8 p_1^3 p_2^3 |\v p_1 - \v p_2|}
\label{eq:divergent triangle}
\end {eqnarray}

There is a very simple way to evaluate the triangle integral
(\ref{eq:triangle}).  First shift $\v l\to \v l+\v p_3$.  Then simply change
variables again by a conformal inversion:
\begin {equation}
   \v l \to {\v l \over l^2} \,,
\end {equation}
and rewrite $\v p_1{-}\v p_3$ and $\v p_2{-}\v p_3$ in terms of
\begin {equation}
   \v q_1 = {\v p_1-\v p_3 \over (\v p_1 - \v p_3)^2} \,,
   \qquad
   \v q_2 = {\v p_2-\v p_3 \over (\v p_2 - \v p_3)^2} \,.
\end {equation}
The result is proportional to the bubble integral (\ref{eq:bubble}),
which is trivial to evaluate.

The same technique can be used for the integral
(\ref{eq:divergent triangle}), which is easiest to implement by
instead considering the convergent integral
\begin {equation}
 \int_l \left[
    {1\over l^4(\v l-\v p_1)^2(\v l-\v p_2)^2} - {1\over l^4 p_1^2 p_2^2}
 \right] .
\end {equation}
The inversion relates this to an integral of the form
\begin {equation}
 \int_l \left[
    {l^2\over (\v l- \v p_1)^2 (\v l-\v p_2)^2} - {1\over l^2}
    \right]
  = {\v p_1 \cdot \v p_2 \over 8 z_{12}}
  \,,
\end {equation}
which is easy to evaluate.

The integral (\ref{eq:x integral}) can be reduced to the others by
rewriting the numerator in terms of denominators as
\begin {equation}
   1 = {1\over2 \v p_1 \cdot \v p_2} \left[
            l^2 + (\v l - \v p_1 - \v p_2)^2
            - (\v l - \v p_1)^2 - (\v l - \v p_2)^2
       \right] .
\end {equation}
It is the only specific case we need of the more general result
\begin {eqnarray}
 \int_l {1\over(\v l-\v p_1)^2(\v l-\v p_2)^2(\v l-\v p_3)^2(\v l-\v p_4)^2}
  &&=
\nonumber\\
    {1\over8 (z_{12}z_{34} + z_{13}z_{24} + z_{14}z_{23})}
    \Biggl[
       {1\over z_{12}z_{13}z_{14}}
&&
       + {1\over z_{21}z_{23}z_{24}}
       + {1\over z_{31}z_{32}z_{34}}
       + {1\over z_{41}z_{42}z_{43}}
    \Biggr] .
\end {eqnarray}


\subsection{Angular averages}
\label{apndx:angular averages}
\begin {eqnarray}
   \left\langle {1\over|\v p + \v q|} \right\rangle_\theta &=& {1\over p_>}
\\
   \left\langle {\cos\theta \over|\v p + \v q|} \right\rangle_\theta
            &=& - {p_< \over 3 p_>^2} \,,
\\
   \left\langle {1\over\cos\theta}
        \left({1\over|\v p - \v q|}  - {1\over|\v p + \v q|}\right)
        \right\rangle_\theta
   &=& {2\over\sqrt{p^2+q^2}} \, {\rm Sinh}^{-1}{p_<\over p_>}
\end {eqnarray}


\subsection{Feynman parameter integrals needed for SU(2) scalar loops}
\label{apndx:Feyn param}

\begin {equation}
    {\cal F}_\alpha[f(x,y,z)] \equiv
    \int_0^1 dx\, dy\, dz
         {f(x,y,z) \,\delta(1-x-y-z) \over
           (yz w_1^2 + zx w_2^2 + xy w_3^2)^\alpha}
\end {equation}

The $w_i$ are positive scalars.

\begin {eqnarray}
   {\cal F}_{3/2}[1]     &=& {2\pi \over w_1 w_2 w_3}
\\
   {\cal F}_{3/2}[x]     &=& {2\pi \over w_2 w_3 (w_1 + w_2 + w_3)}
\\
   {\cal F}_{3/2}[x^2]   &=& {\pi \over w_2 w_3} \left[
                               {1 \over (w_1 + w_2 + w_3)}
                               + {w_1 \over (w_1 + w_2 + w_3)^2}
                             \right]
\\
   {\cal F}_{3/2}[xy]    &=& {\pi \over w_3 (w_1 + w_2 + w_3)^2}
\\
   {\cal F}_{3/2}[x^3]   &=& {\pi \over 4 w_2 w_3} \left[
                               {3 \over (w_1 + w_2 + w_3)}
                               + {3 w_1 \over (w_1 + w_2 + w_3)^2}
                               + {2 w_1^2 \over (w_1 + w_2 + w_3)^3}
                             \right]
\\
   {\cal F}_{3/2}[x^2 y] &=& {\pi \over 4 w_3} \left[
                               {1 \over (w_1 + w_2 + w_3)^2}
                               + {2 w_1 \over (w_1 + w_2 + w_3)^3}
                             \right]
\\
   {\cal F}_{3/2}[xyz]   &=& {\pi \over 2 (w_1 + w_2 + w_3)^3}
\\
   {\cal F}_{1/2}[1]     &=& {\pi \over w_1 + w_2 + w_3}
\\
   {\cal F}_{1/2}[x]     &=& {\pi \over 4} \left[
                               {1 \over (w_1 + w_2 + w_3)}
                               + {w_1 \over (w_1 + w_2 + w_3)^2}
                             \right]
\end {eqnarray}
The ${\cal F}_{3/2}$ results can be obtained from the
${\cal F}_{1/2}$ results by differentiating with respect to the $w_i$ and
using $x+y+z=1$.


\subsection{Scalar integrals needed for SU(2)}
\label{apndx:bleah}

\begin {eqnarray}
   \v J &\equiv& z_{12}\v p_3 + z_{23}\v p_1 + z_{31}\v p_2
\\
   \v K &\equiv& \v p_1 + \v p_2 + \v p_3
\end {eqnarray}
\begin {eqnarray}
&&
 \int_l {l_i\over(\v l-\v p_1)^2(\v l-\v p_2)^2(\v l-\v p_3)^2} =
    {J_i\over8 z_{12}z_{23}z_{31}(z_{12}+z_{23}+z_{31})}
\\
&&
 \int_l {l_i l_j\over(\v l-\v p_1)^2(\v l-\v p_2)^2(\v l-\v p_3)^2} =
    {\delta_{ij}\over16(z_{12}+z_{23}+z_{31})}
    + {J_i J_j\over16z_{12}z_{23}z_{31}(z_{12}+z_{23}+z_{31})^2}
\nonumber\\ && \qquad\qquad\qquad\qquad\qquad\qquad\qquad\qquad
    + {z_{12} p_{3i} p_{3j} + z_{23} p_{1i} p_{1j}
          + z_{31} p_{2i} p_{2j}
      \over16z_{12}z_{23}z_{31}(z_{12}+z_{23}+z_{31})}
\\
&&
 \int_l {l_i l_j l_k\over(\v l-\v p_1)^2(\v l-\v p_2)^2(\v l-\v p_3)^2} =
    - {J_i\delta_{jk} + J_j\delta_{ki} + J_k\delta_{ij}
            \over64(z_{12}+z_{23}+z_{31})^2}
    - {K_i\delta_{jk} + K_j\delta_{ki} + K_k\delta_{ij}
            \over64(z_{12}+z_{23}+z_{31})}
\nonumber\\ && \qquad\qquad
    - {J_i J_j J_k
            \over32z_{12}z_{23}z_{31}(z_{12}+z_{23}+z_{31})^3}
    - { \langle 9 z_{12}^2 p_{3i} p_{3j} p_{3k}
           + 18 z_{12} z_{13} p_{2i} p_{2j} p_{3k} \rangle
        \over 64 z_{12}z_{23}z_{31}(z_{12}+z_{23}+z_{31})^2}
\nonumber\\ && \qquad\qquad
    - { \langle 9 z_{12} p_{3i} p_{3j} p_{3k} \rangle
        \over 64 z_{12}z_{23}z_{31}(z_{12}+z_{23}+z_{31})}
\end {eqnarray}
The angle brackets in the last expression denote averaging over all
permutations of $(i,j,k)$ and all permutations of $(\v p_1,\v p_2, \v p_2)$.


\section {Summary of diagrams}
\label{apndx:summary}

\subsection {U(1) case: Feynman gauge}
\label{apndx:feynman gauge}

\begin {eqnarray}
\dlam^{\rm(a,c,h,\counterb)}
   &=& \hbox{eq.~(\ref{eq:div3})}
\nonumber\\
   &=& N g^8 \int_{pq}
       \left[ 3 f_p f_q^3 + (3-2C_\lambda)^2 f_p^2 f_q^2 + 3 f_p^3 f_q \right]
       \left(-{6\over\pi^2\Lambda}\right)
\nonumber\\
   &=& N g^8 \intFF \left(- {6\over\pi^2\Lambda}\right)
       + N g^8 \int_{pq} f_p^3 f_q \left(-{24\over\pi^2\Lambda}\right)
\label{eq:B1}
\\
\dlam^{\rm(b,g,\countera)}
   &=& \hbox{eq.~(\ref{eq:div2})}
\nonumber\\
   &=& N g^8 \int_{pq}
       \left[ f_p f_q^3 + (3-2C_\lambda) f_p^2 f_q^2 + 3 f_p^3 f_q \right]
       \left({6\over q} + {12\over\pi^2\Lambda}\right)
\nonumber\\
   &=& N g^8 \intFF \left[ {1\over p_>} (3 + 3 x^{-1})
            + {12\over\pi^2\Lambda} \right]
       + N g^8 \int_{pq} f_p^3 f_q
              \left({12\over q} + {24\over\pi^2\Lambda}\right)
\label{eq:B2}
\\
\dlam^{\rm(f)}
   &=& \hbox{eq.~(\ref{eq:div1})}
\nonumber\\
   &=& N g^8 \intFF \Biggl[
          -{6\over p} - {6\over q} - {6\over|\v p+\v q|}
          + {3 \cos\theta\over 2|\v p+\v q|}
          + {3(p^2+q^2)\over pq|\v p+\v q|}
          - {6\over\pi^2\Lambda}
       \Biggr]
\nonumber\\
   &=& N g^8 \intFF
       \left[
          {1\over p_>}
              \left( -12 - 3 x^{-1} + {\textstyle{5\over2}} x \right)
          - {6\over\pi^2\Lambda}
       \right]
\label{eq:B3}
\\
\dlam^{\rm(i)}
   &=& \hbox{eq.~(\ref{eq:Vg})}
\nonumber\\
   &=& N g^8 \intFF {1\over p_>} (-9)
\\
\dlam^{\rm(j)}
   &=& \hbox{eqs.~(\ref{eq:Vh0}, \ref{eq:Vh})}
\nonumber\\
   &=& N g^8 \intFF {1\over p_>}
       (18-4x)
\\
\dlam^{\rm(k)}
   &=& - 6 N g^8 \intFF \int_l
    { (2\v l-\v p)\cdot(2\v l-\v p+2\v q) \,
      (2\v l+\v q)\cdot(2\v l-2\v p+\v q)
      \over
      l^2 (\v l-\v p)^2(\v l+\v q)^2(\v l-\v p+\v q)^2
    }
\nonumber\\
   &=& N g^8 \intFF \int_l
    \Biggl[
      -{6\over p} -{6\over q}
      +{9(p^2 + q^2)\over 2pq}
          \left({1\over|\v p+\v q|} + {1\over|\v p-\v q|}\right)
\nonumber\\ && \qquad\qquad\qquad\qquad
      + \left( {3(2p^4+5p^2q^2+2q^4)\over 4 pq \v p\cdot\v q}
             - {3 \v p \cdot \v q \over pq} \right) 
          \left({1\over|\v p+\v q|} - {1\over|\v p-\v q|}\right)
    \Biggr]
\nonumber\\
   &=& N g^8 \intFF {1\over p_>}
       \biggl[
         -6 + 3 x^{-1} + 7 x
\nonumber\\ && \qquad\qquad\qquad\qquad
         -\left({\textstyle {15\over2} + 3 x^{-2} + 3 x^2}\right)
               (1+x^2)^{-1/2}\, {\rm Sinh}^{-1}x
       \biggr]
\end {eqnarray}
The graphs of figs.~\ref{fig:NLO}d and e and the second term in the mass
counter-term \ref{eq:mbare} have been ignored above, as they trivially
cancel each other.

The infrared divergences of (\ref{eq:B1}) through (\ref{eq:B3}) were regulated
with dimensional regularization, but, for the sake of making cancelations
explicit, we have put the linear divergences back in by hand by writing
\begin {equation}
   \int_l {1\over l^4} = {1\over2\pi^2\Lambda} \,,
\end {equation}
where $\Lambda$ is an infrared momentum cut-off.


\subsection{Abelian graphs: Landau gauge}
\label{apndx:Landau}

\begin {equation}
   s(x) \equiv \cases{
                   1, & \hbox{U(1) theory} \cr
                   x, & \hbox{SU(2) theory}\cr
               }
\end {equation}

\begin {eqnarray}
\dlam^{\rm(a{-}d,g,h,\countera,\counterb)}
   &=& s\left({\textstyle{9\over128}}\right) N g^8 \int_{pq}
        f_p^3 f_q \, {12\over q}
\\
\dlam^{\rm(f)}
   &=& s\left({\textstyle{9\over128}}\right) N g^8 \intFF
          {1\over p_>}
              \left( -8 + 4 x - {\textstyle{4\over5}} x^2 \right)
\\
\dlam^{\rm(i)}
   &=& s\left({\textstyle{3\over128}}\right) N g^8 \intFF {1\over p_>}
          \left( -4 - {\textstyle{2\over5}} x^2  \right)
\\
\dlam^{\rm(j)}
   &=& s\left({\textstyle{3\over128}}\right) N g^8 \intFF {1\over p_>}
          \left( 8 - 4 x + {\textstyle{4\over5}} x^2  \right)
\\
\dlam^{\rm(k)}
   &=& s\left({-\textstyle{3\over128}}\right) N g^8 \intFF {1\over p_>}
       \biggl[
         -2 + 3 x^{-1} + {\textstyle{11\over2}} x
         + {\textstyle{2\over5}} x^2
\nonumber\\ && \qquad\qquad\qquad\qquad
         -\left({\textstyle {15\over2} + 3 x^{-2} + 3 x^2}\right)
               (1+x^2)^{-1/2}\, {\rm Sinh}^{-1}x
       \biggr]
\end {eqnarray}

For non-Abelian graphs, see eqs. (\ref{eq:dlam m}),
(\ref{eq:dlam n}) and (\ref{eq:dlam jkl}).


\end {document}